\documentclass{aastex}
\usepackage{spr-astr-addons}
\usepackage{url}
\usepackage{psfig}   
\usepackage{subfigure}

\RequirePackage{color}

\begin{document}

\title{Landau level-superfluid modified factor and effective X/$\gamma$-ray coefficient of a magnetar }
\slugcomment{Not to appear in Nonlearned J., 45.}
\shorttitle{Two important parameters of a magnetar }
\shortauthors{Z. F. Gao et al.}
\author{Z. F. Gao\altaffilmark{1,2,3}}
\altaffiltext{1}{Xinjiang Astronomical Observatory, CAS, 40-5 South Beijing Road, Urumqi Xinjiang, 830011, China zhifu$_{-}$gao@uao.ac.cn}
\altaffiltext{2}{Key Laboratory of Radio Astronomy, Chinese Academy of Sciences£¬Nanjing, 210008, China}
\altaffiltext{3}{Graduate University of the Chinese Academy of Scienes, 19A Yuquan Road, Beijing, 100049, China}
\author{Q. H. Wang \altaffilmark{4}}
\affil{Department of  Astronomy, Nanjing University, Nanjing, 2100093, China}
\author{ N. Wang \altaffilmark{1,2}}
\altaffiltext{1}{Xinjiang Astronomical Observatory, CAS, 40-5 South Beijing Road, Urumqi Xinjiang, 830011, China}
\altaffiltext{2}{Key Laboratory of Radio Astronomy, Chinese Academy of Sciences£¬Nanjing, 210008, China}
\author{C. K. Chou \altaffilmark{5}}
\affil{National Astronomical Observatories, Chinese Academy of Sciences, Beijing, 10012, China}
\author{W. S. Huo \altaffilmark{6,7}}
\altaffiltext{6}{School of Physics, Xinjiang University, Urumqi Xinjiang, 830011, China}
\altaffiltext{7}{ Xinjiang University-National Astronomical Observatories Joint Center for Astrophysics,Urumqi Xinjiang, 830011, China}
\begin{abstract}

As soon as the energy of electrons near the Fermi surface
are higher than $Q$, the threshold energy of inverse $\beta-$ decay,
the electron capture process will dominate. The resulting high-energy
neutrons will destroy anisotropic ${}^3P_2$ neutron superfluid Cooper
pairs.  By colliding with the neutrons produced in the process $n+
(n\uparrow n\downarrow)\longrightarrow n+ n+ n$, the kinetic energy of
the outgoing neutrons will be transformed into thermal energy. The
transformed thermal energy would transported from the star interior to
the star surface by conduction, then would be transformed into radiation
energy as soft X-rays and gamma-rays. After a highly efficient
modulation within the pulsar magnetosphere, the surface thermal
emission (mainly soft X/$\gamma$-ray emission) has been shaped into a
spectrum with the observed characteristics of magnetars. By introducing
two important parameters: Landau level-superfluid modified factor
and effective X/$\gamma$-ray coefficient, we numerically simulate the
process of magnetar cooling and magnetic field decay, and then  compute
magnetars' soft X/$\gamma$-ray  luminosities $L_{X}$. Further, we
obtain aschematic diagrams of $L_{X}$ as a function of magnetic field
strength $B$. The observations are compared with the calculations.
\end{abstract}


\section{Introduction}

Magnetars are ultra-magnetized neutron stars (NSs) with
magnetic fields largely in excess of the quantum critical field
$B_{\rm cr} = m^{2} _{e}c^{3}/e\hbar$ = 4.414 $\times10^{13}$ G,
at which the energy between Landau levels of electrons equals the
rest-mass energy of a electron \citep{Duncan92, Thompson95, Thompson96}.
Unlike ordinary radio pulsars, powered by their rotational energy loss, or shining in
X-rays thanks to the accretion of matter from their companion stars, magnetars¡¯ persistent X-ray luminosities, are
instead believed to be powered by the decay of their exceptionally strong magnetic fields \citep{Colpi00,Thompson95,
Thompson96,Woods04, Mereghetti08}.

The majority of magnetars are classified into two NS
populations historically that were independently discovered
through different manifestations of their high-energy emission \citep{Colpi00,
Kouveliotou98,Woods04}: the soft gamma-ray repeaters (SGRs),
which give sporadic bursts of hard X-rays/soft$\gamma$-rays as
well as rare, very luminous ($\sim 10^44$ erg~$s^-1$) ¡®giant flares¡¯,
and the anomalous X-ray pulsars (AXPs), so named due
to their high X-ray luminosities ($\sim 10^{34}- 10^{36}$ erg~$s^-1$) and
unusually fast spin-down rates, with no evidence of variation due to binary motion, which are distinct from both
accreting X-ray binaries and isolated radio pulsars. Both
AXPs and SGRs have common properties: stochastic outbursts (lasting from days to years) during which they emit
very short X/$\gamma$-ray bursts; rotational periods in a narrow
range $P\sim (6\sim 12)$ s; compared to other isolated neutron
stars, large period derivatives of $(\sim 10^{-13}- 10^{-10})$ s~$s^-1$
; rather soft X-ray spectra ($kT < 10$  keV) that can be fitted by
the sum of a blackbody model with a temperature $kT \sim $
0.5 keV and a power-law tail with photon index $\sim 3 \sim 4$
\citep{Mereghetti02} and, in some cases, associated with
supernova remnants (SNRs)\citep{Duncan92, Mereghetti08}. In a few AXPs, good fits are obtained
equivalently with two blackbodies \citep{Halpern05} or
other combinations of two spectral components.

With the exception of SGR 0526-66, SGRs tend to have
harder spectra below 10 keV than AXPs, and also suffer of a
larger interstellar absorption, which makes the detection of
blackbody-like components more difficult. For SGR 1806-20 and SGR 1900+14, most of their soft X-ray spectra have
been well fit with power-laws of photon index $\sim$ 2. Nevertheless, when good quality spectra with sufficient statistics are
obtainable, blackbody-like components with $kT\sim$ 0.5 keV
can be detected also in these sources\citep{Mereghetti05,
Mereghetti06a, Mereghetti06b}. These data demonstrate that emissions from
magnetars in the soft X-ray band are predominantly of thermal origin, but the emerging spectrum is far more intricate
than that with a simple Planckian. This is not surprising if we
take into account the presence of a strongly magnetized atmosphere and/or the effects of resonant cyclotron scattering
within the magnetosphere of a magnetar \citep{Tong09}.

Observations from the Rossi X-Ray Timing Explorer
(RXTE) and the International Gamma-Ray Astrophysics
Laboratory (INTEGRAL)have revealed that magnetars are
luminous, persistent sources of 100 keV X-rays \citep{Beloborodov07}.
This high-energy component, as distinct from the soft X-ray component,
has a harder spectrum, and peaks above 100 keV. The luminosity in this band
could be comparable or even exceed the thermal X-ray luminosity from the star
surface. These hard X-rays could be emitted only in the exterior of a magnetar,
which illustrates the existence of an active plasma corona \citep{Beloborodov07}.
However, what we care about is the mechanism for the magnetar soft X-ray/$\gamma$
-ray emission in this article.

To explain the soft X-ray/$\gamma$-ray emission of magnetars, various ideas
have been put forward \citep{Heyl97, Heyl98, Pons06,Rheinhardt03,Thompson96,
Thompson00, Thompson02}. There are also promising physical models that explain
the absence of cyclotron features and the hard X-ray tails \citep{Lyutikov06,
Nobili08}. The thermal emission model has been investigated extensively in the
last decade, and significant progress has been achieved by many authors. However,
some of the most fundamental questions are still unanswered. For example, the
question of where the heat comes from is becoming more and more significant.

In the ¡®fall-back disk¡¯ model, different mechanisms for
the disk formation and the relationship between the disk
and the observed magnetar soft X-ray/$\gamma$-ray luminosity
have been taken into consideration \citep{Alpar01, Chatterjee00, Ghosh97, Marsden01,
Tong10, van95}. The enhanced X-ray emission and the expected optical/IR flux from
magnetars can be interpreted as due to the evolution of disks
after they have been pushed back by the bursts. The transient behavior
of XTEJ1810-197 has been instead explained in terms of a fall-back disk
subject to viscous instability  \citep{Ertan08}. However, the primary flaw of this
model lies in the fact that it is unable to give a clear accounting for the bursts and
flares \citep{Ertan06}. As a result, other possible mechanisms for the magnetar soft
X-ray/$\gamma$-ray emission have to be added.

The origin of magnetars is another interesting and important issue.
A currently popular hypothesis is that magnetars are formed from rotating
proto-neutron stars, and rapid differential rotation and convection would
result in an efficient $\alpha -\Omega$ dynamo \citep{Duncan92,Duncan96,
Thompson93, Vink06}. In the context of the $\alpha -\Omega$ dynamo model,
a mechanism (called a twisting magnetosphere) has been proposed to describe
the radiative properties of magnetars \citep{Thompson00}. According
to the ¡®twisting magnetosphere¡¯ model, the energy caught in
the twisting magnetic field gradually dissipates into X-rays.
For a magnetar, the persistent soft X-ray emission can be induced by the twisting
of the external magnetic field, while the persistent hard X-ray emission originates
in a transition layer between the corona and the atmosphere \citep{Beloborodov07, Thompson00}.
The gradual dissipation of the magnetospheric currents can produce the
persistent soft $\gamma$-ray emission \citep{Thompson05}. In some sources (primarily AXPs),
it is very difficult to evaluate the bolometric losses by a steep power-law
component in the X-ray spectrum, which is not observed
below $\sim $ 0.5 keV because of absorption by the star¡¯s inner
matter \citep{Thompson02}. Unfortunately, we have not so far found any evidence
supporting the idea that magnetars are formed from rotating proto-neutron stars.
There is as yet no mechanism to explain such a high efficiency of energy transformation
in the $\alpha -\Omega$ model. Moreover, when investigating solar flare using this model,
apart from the difficulty in explaining such a high energy transformation efficiency, there
are also many other problems to be settled at present. According to the above analysis, the
$\alpha -\Omega$ dynamo is still just a hypothesis \citep{Gao11b, Hurley05,
Mereghetti08, Vink06}.

In this paper, we focus on the interior of a magnetar, where the
reaction of electron capture (EC) $e^{-}+ p\rightarrow n + \nu_{e}$
proceeds.  As we know, EC, also called `the inverse $\beta$-decay', is
a key physical process for nucleosynthesis and neutrino production in
supernova (SN), especially for core-collapsed SN (including type SNII,
SNIb and SNIc)\citep{Langanke00,Luo06,Peng01}. It not only carries away
energy and entropy in the form of neutrinos, but also reduces the number
of electrons in the interior of a SN.  The ways of calculating parameters
relating to EC are dependent on models.  For example, specific but
representative parameters encountered during the initial stages of core
formation in a SN are temperatures $T\simeq 3\times 10^{10}$ K($kT\simeq$
2.4 MeV), densities $\rho\simeq1.4\times10^{12}$ g cm$^{-3}$, electron
Fermi energies $E_{F}(e)\simeq$ 35 MeV, and the Fermi energies of
neutrinos produced in the process of EC, $E_{F}(\nu)\simeq$ 25 MeV
\citep{Lamb76,Mazurek76}. Unlike any other way of dealing with EC, by
introducing related parameters and comparing our results with the
observations, we numerically simulate the whole EC process accompanied
by decay of the magnetic field  and fall of the internal
temperature.  According to our model, superhigh magnetic fields of
magnetars could be from the induced magnetic fields at a moderate
lower temperature due to the existence of ${}^3P_2$ anisotropic neutron
superfluid, and the maximum of induced magnetic field is estimated
to be (3.0$\sim$ 4.0) $\times 10^{15}$ G \citep{Peng07,Peng09}.  In
the interior of a magnetar(mainly in the outer core), superhigh
magnetic fields give rise to the increase of the electron Fermi energy
\citep{Gao11a}, which will induce EC (if the energy of a electron is
higher than the value of $Q$, the threshold energy of inverse $\beta-$ decay).
The resulting high-energy  neutrons will destroy anisotropic ${}^3P_2$
neutron Cooper pairs, then the ${}^3P_2$ anisotropic superfluid and the
superhigh magnetic field induced by the ${}^3P_2$ Cooper pairs will
disappear. By colliding with the neutrons produced in the process $n+
(n\uparrow n\downarrow)\longrightarrow n+ n+ n$, the kinetic energy of
the outgoing neutrons will be transformed into thermal energy. The
transformed thermal energy would be transported from the star interior to
the star surface by conduction, then would be transformed into radiation
energy as soft X-rays and gamma-rays.  For further details, see
Sec. 3.

In order to be feasible for our model, two factors must be
taken into account. The first is that too much energy is lost in
the process of thermal energy transportation, due to the inner matter'
absorption  and the emission of neutrinos escaping from the interior of a
magnetar. The second is that most of thermal energy transported to the
star surface is carried away by the surface neutrino flux, only a small
fraction  can be converted into radiation energy as soft X/$\gamma$-rays.
Taking into account of the above-mentioned two factors, we introduce the
energy conversion coefficient $\epsilon$ and the thermal energy
transportation coefficient $\theta$ in the latter calculations.

The remainder of this paper is organized as follows: in
Sect.2we calculate the electron Fermi energy $E_{F}(e)$, the
average kinetic energy of the outgoing EC neutrons $\langle E_{n}\rangle$,
and the average kinetic energy of the outgoing EC neutrinos $\langle E_
{\nu}\rangle$. In Sect. 3, we describe our interpretation of soft
X/$\gamma$-ray emission of magnetars, compute the luminosity $L_{X}$ and
obtain a diagram of $L_{X}$ as a function of $B$. A brief conclusion is given
in Sect.4, and a special solution of the electron
Fermi energy is derived briefly in Appendix.

\section{The calculations of $E_{F}(e)$, $\langle E_{n}\rangle$ and $\langle E_{\nu}\rangle$ }
\subsection{Electron Fermi energy and energy state densities of particles}
In the presence of an extremely strong magnetic field ($B\gg B_{cr}$), the Landau column becomes a very long
and very narrow cylinder along the magnetic field, the electron Fermi
energy is determined by
\begin{eqnarray}
&&\frac{3\pi}{B^{*}}(\frac{m_{e}c}{h})^{3}(\gamma_{e})^{4}\int_{0}^{1}
(1-\frac{1}{\gamma^{2}_{e}}-\chi ^{2})^{\frac{3}{2}}d\chi\nonumber\\
&&-2\pi\gamma_{e}(\frac{m_{e}c}{h})^{3}\sqrt{2B{*}}=N_{A}\rho Y_{e},
\end{eqnarray}
where $B^{*}$, $\chi$ and $\gamma_{e}$ are three non-dimensional
variables, which are defined as ${B^{*}=B/B_{cr}}$, $\chi =(\frac{p_
{z}}{m_{e}c})/(\frac{E_{F}}{m_{e}c^{2}})= p_{z}c/E_{F}$ and $\gamma_{e}
= E_{F}(e)/m_{e}c^{2}$, respectively; $1/\gamma^{2}_{e}$ is  the
modification factor; $N_{A}=$ 6.02$\times 10^{23}$ is the Avogadro
constant; $Y_{e}=Y_{p}= Z/A$, here $Y_{e}$, $Y_{p}$, $Z$ and $A$ are
the electron fraction, the proton fraction, the proton number and
nucleon number of a given nucleus, respectively \citep{Gao11a}.  From
Eq.(1), we obtain the diagrams of $E_{F}(e)$ vs. $B$,  shown as in Fig. 1.

 \begin{figure}[th]
\centering
  \vspace{0.5cm}
  \includegraphics[width=8.2cm,angle=360]{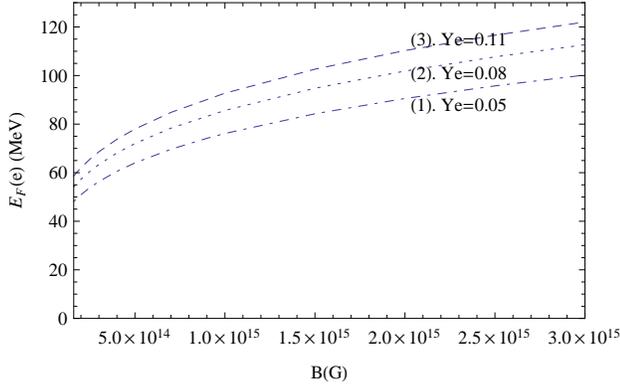}
  \caption{The relation of $E_{F}(e)$ and $B$. The range of
$B$ is (1.6$\times 10^{14}\sim$ 3.0$\times 10^{15}$)G, and $\rho$
=2.8$\times 10^{14}$ g~cm$^{-3}$. Dot-dashed line, dotted line and dashed
line  for $Y_{e}$=0.05, $Y_{e}$=0.08 and
$Y_{e}$=0.11, respectively.}
  \label{ 3:fig}
\end{figure}

From Fig. 1, it's obvious that $E_{F}(e)$ increases
with the increase in $B$ when $\rho$ and $Y_{e}$ are given.
The high $E_{F}(e)$ could be generated by the release of magnetic field energy.
According to Eq.(11.5.2) of Page 316 in Shapiro \& Teukolsky (1983), in
empty space for each species $j$ we would have
\begin{equation}
\rho_{j}dE_{j} = d^{3}n_{j} = V_{1}\frac{d^{3}p_{j}}{h^{3}}g_{j},
\end{equation}
where $V_{1}$ denotes the normalization volume, $n_{j}$ is
the particle number of species $j$, $\rho_{j}$ is the energy
state density for species $j$ (the particle number per unit energy),
and $g_{j}$ is the degeneracy for species $j$ ($g_{j}$ = 2 for fermions).
In the vicinity of a Fermi surface, $E_{j}\sim E_{F}(j)$ ($j = e, p, n,
\nu$).  Since protons and neutrons are degenerate and nonrelativistic,
we get the approximate relations $E_{p}\sim E_{F}^{'}(p)= p_{p}^{2}/2m_{p}$ and  $E_{n}\sim E_{F}
^{'}(n)=p_{n}^{2}/2m_{n}$, then we obtain $dE_{p} = p_{p}dp_{p}/m_{p}$ and
$dE_{n}= p_{n}dp_{n}/m_{n}$, where $E_{F}^{'}(n)$ and $E_{F}^{'}(p)$ are
the neutron Fermi kinetic energy and the proton Fermi kinetic energy,
respectively \citep{Shapiro83}.   Since $\rho_{p}dE_{p}= 2 \times \frac
{4\pi}{h^{3}}p_{p}^{2}dp_{p}=\frac{8\sqrt{2}\pi}{h^{3}}m_{p}^{3/2}E_{p}^{1/2}
dE_{p}$ and $\rho_{n}dE_{n} = 2 \times \frac{4\pi}{h^{3}}p_{n}^{2}dp_{n}
=\frac{8\sqrt{2}\pi}{h^{3}}m_{n}^{3/2}E_{n}^{1/2}dE_{n}$, the energy state
densities of neutrons and protons can be approximately written as

\begin{equation}
\rho_{n}\simeq \frac{8\sqrt{2}\pi}{h^{3}}(m_{n})^{\frac{3}{2}}E_{F}(n)^{\frac{1}{2}},
\end{equation}
and
\begin{equation}
\rho_{p}\simeq \frac{8\sqrt{2}\pi}{h^{3}}(m_{p})^{\frac{3}{2}}E_{F}(p)^{\frac{1}{2}},
\end{equation}
respectively. If we define the electron Fermi momentum $p_{F}(e)$ by
\begin{equation}
p_{F}(e)= [E_{F}^{2}(e)-m_{e}^{2}c^{4}]^{\frac{1}{2}}/c,
\end{equation}
then we find that the electron number in a unit volume is
\begin{equation}
n_{e}= \frac{2}{h^{3}}\int_{0}^{p_{F}(e)}4\pi p_{e}^{2}dp_{e}=\frac{8\pi}{3h^{3}}p_{F}^{3}(e).
\end{equation}
The electron energy density $\rho_{e}$ is determined by
\begin{equation}
\rho_{e}= \frac{4\pi p_{e}^{2}}{h^{3}}\frac{dp_{e}}{dE_{e}}=\frac{4\pi p_{e}E_{e}}{c^{2}h^{3}}.
\end{equation}
Since the neutrino is massless, by energy conservation $E_{\nu} = E_{e}-Q
= p_{\nu}c$  in the process of EC, we obtain the expression of the neutrino
energy state density $\rho_{\nu}$
\begin{equation}
\rho_{\nu}=\frac{(E_{e}-Q)^{2}}{2\pi^{2}\hbar^{3}c^{3}},
\end{equation}
where $p_{\nu}$ is the neutrino momentum.

\subsection{The calculations of $E_{F}(e)$, $\langle E_{n}\rangle$ and $\langle E_{\nu}\rangle$ }
We focus on non-relativistic, degenerate nuclear matter and ultra-
relativistic, degenerate electrons under $\beta$-equilibrium implying
the following relationship among chemical potentials (called the Fermi
energies $E_F$)of the particles: $\mu_p+\mu_e=\mu_n$, where the neutrino
chemical potential is ignored.   In the case of $0.5\rho_{0}\leq \rho
\leq  2\rho_{0}$, the following expressions are hold approximately:
$ E_{F}^{'}(n)=60(\rho/\rho_{0})^{\frac{2}{3}}$ MeV, $E_{F}^{'}(p)=1.9
(\rho/\rho_{0})^{\frac{4}{3}}$ MeV, $(m_n-m_p)c^{2}$= 1.29 MeV (c.f.
Chapter $11$ of (Shapirol \& Teukolsky, 1983), where $\rho_{0}$=2.8$
\times 10^{14}$ g~cm$^{-3}$ is the standard nuclear dencity.  In this
work, for convenience, we set $\rho=\rho_{0}$, yielding the threshold
energy of electron capture reaction $Q =E_F(n)-E_F(p)=(m_n-m_p)c^{2}
+(p^{2}_{F}(n)/2m_n- p^{2}_{F}(p)/2m_p)$=1.29 MeV+(60-1.9)MeV =59.39 MeV,
then the range of $E_{e}$ is (59.39 MeV $\sim E_{F}(e)$).  For an
outgoing neutrons the value of $E_{k}(n)$ is not less than that of $E_{F}
^{'}(n)$, otherwise the process of $e^{-}+p \rightarrow n+\nu_e$ will not
occur, so the range of $E_{k}(n)$ is assumed to be ($E_{F}^{'}(n)\sim
\langle E_{n}\rangle$).  By employing energy conservation via $E_{\nu}+
E_{n}=E_{e}+E_{p}$, the Fermi energies of neutrinos resulting in the
process of EC, $E_{F}(\nu)$, can be calculated by
\begin{equation}
E_{F}(\nu)=E_{F}(e)-Q=E_{F}(e)-59.39 \rm MeV.
\end{equation}
The electron capture rate $\Gamma$ is defined as the number of
electrons captured by one proton per second, and can be computed
by using the standard charged-current $\beta$-decay theory.  The
expression for $d\Gamma$ reads:
\begin{eqnarray}
 &&d\Gamma = \frac{2\pi}{\hbar}G_{F}^{2}C_{V}^{2}(1+3a^{2})\nonumber\\
&&(1-f_{\nu})\rho_{\nu}dE_{\nu}\delta(E_{\nu}+Q-E_{e}),
\end{eqnarray}
where $G_{F}$=1.4358$\times10^{-49}$ erg~cm$^{3}$, is the universal
Fermi coupling constant in the Weinberg-Salam-Glashow theory; $C_{V}$= 0.9737
is the vector coupling constant; $a$ is the ratio of the
coupling constant of the axial vector weak interaction constant
to that of the  vector weak interaction, $a$ = 1.253 experimentally;
the quantity $1-f_{\nu}$ is the neutrino `blocking factor'giving the
fraction of unoccupied phase space for neutrinos.  For both electrons
and neutrinos we shall assume Fermi-Dirac equilibrium distributions,
\begin{equation}
f(j)=\frac{1}{exp[(E_{j}-\mu_{j})/kT)]+ 1}
 \end{equation}
To find the total electron capture rate per
proton, we integrate over all initial electron states and over $dE_{\nu}$
\begin{eqnarray}
&&\Gamma=\frac{2\pi}{\hbar}G_{F}^{2}C_{V}^{2}(1+3a^{2})\int_{Q}^{E_{F}(e)}f_{e}(1-f_{\nu})\nonumber\\
&&\times \rho_{e}dE_{e}\rho_{\nu}dE_{\nu}\delta(E_{\nu}+Q-E_{e})\nonumber\\,
&&=\frac{2\pi}{\hbar}\frac{G_{F}^{2}C_{V}^{2}(1+3a^{2})}{(2\pi^{2}\hbar^{3}c^{3})^{2}}\int_{Q}^{E_{F}(e)}E_{e}^{2}-m^{2}_{e}c^{4})^{\frac{1}{2}}\nonumber\\
&&\times (E_{e}(E_{e}-Q)^{2}f_{e}(1-f_{\nu})dE_{e},
\end{eqnarray}
where  Eq.(7) for the electron energy state density is used. In the interior of a NS, for neutrinos (antineutrinos),
 $(1-f_{\nu})$= 1; for electrons, when $E_{e} < E_{F}(e)$, $f_{e}$= 1, and when $E_{e}
> E_{F}(e)$, $f_{e}$ = 0.  The average kinetic energy of the outgoing neutrinos
$\langle E_{\nu}\rangle$ can be calculated by
\begin{eqnarray}
&&\langle E_\nu \rangle= \int E_\nu d\Gamma /\Gamma =\int_{Q}^{E_{F}(e)}\nonumber\\
&&(E_{e}^{2}-m^{2}_{e}c^{4})^{\frac{1}{2}}E_{e}(E_{e}-Q)^{3}f_{e}(1-f_{\nu})dE_{e}/I,
\end{eqnarray}
where $I=\int_{Q}^{E_{F}(e)}(E_{e}-Q)^{2}E_{e}(E_{e}^{2}-m_{e}^{2}c^{4})^{\frac
{1}{2}}f_{e}(1-f_{\nu})dE_{e}$.
The average kinetic energy of the outgoing neutrinos
$\langle E_{\nu} \rangle$ can be estimated by
\begin{equation}
\langle E_{n}\rangle = E_{F}(e)- \langle E_{\nu}\rangle+0.61 \rm MeV.
 \end{equation}
In this paper, for the purpose of convenient calculation, we set
$\rho=\rho_{0}$, $Y_{e}=0.12$ and and $E_{F}(e)$=43.44$(\frac{B}
{B_{cr}})^{\frac{1}{4}}$ MeV, further details are presented in
Appendix.  The range of $B$ is assumed to be (1.5423$\times 10^
{14}\sim$ 3.0$\times 10^{15}$) G, where 1.5423$\times 10^{14}$ G
is the minimum of $B$ denoted as $B_{f}$.  When $B$ drops below
$B_{f}$, the direct Urca process is quenched everywhere in the
magnetar interior.  If we want to get the value of $L_{X}$ in
any superhigh magnetic field, the quantities $E_{F}(e)$, $E_{F}
(\nu)$, $\langle E_{n}\rangle$ and $\langle E_{\nu}\rangle$ must
firstly be computed.   The calculation results are shown in Table 1.

\begin{table}
\small
\caption{The calculated  values of $E_{F}(e), \langle E_{\nu} \rangle$
  and $\langle E_{n}\rangle$\label{tb1-4}.}
\begin{tabular}{@{}crrrr@{}}
\tableline
 B       & \multicolumn{1}{c}{$E_{F}(e)$\tablenotemark{a}}  &   $E_{F}(\nu)$ & \multicolumn{1}{c}{$\langle E_{\nu}\rangle$\tablenotemark{b}}
 &\multicolumn{1}{c}{$\langle E_{n}\rangle$\tablenotemark{c}}\\
 (G)       &(MeV)      &(MeV)     & (MeV)       &(MeV)\\
\tableline
1.6$\times 10^{14}$&59.94   & 0.55    & 0.41  &60.14   \\
2.0$\times 10^{14}$ &63.38  &3.99   &3.01    &60.98   \\
2.5$\times 10^{14}$&67.01   & 7.62    & 5.78 &61.84   \\
3.0$\times 10^{14}$ &70.14 &10.75  &8.19    &62.56   \\
4.0$\times 10^{14}$ &75.37  &15.98    &12.48   &63.50  \\
5.0$\times 10^{14}$ &79.69  &20.30  &15.63   &64.67 \\
7.0$\times 10^{14}$&86.69   &27.30   &21.15  &66.15     \\
9.0$\times 10^{14}$&92.31   &32.92  &25.62  &67.30     \\
1.0$\times 10^{15}$&94.77    &35.38  &27.58  &67.80    \\
1.5$\times 10^{15}$&104.88    &45.49  &35.70 & 69.79  \\
2.0$\times 10^{15}$&112.70 &53.31  &42.01  &71.30    \\
2.5$\times 10^{15}$&119.17 &59.78 &47.25  &72.53    \\
2.8$\times 10^{15}$& 122.59 &63.20 &50.03 &73.17  \\
3.0$\times 10^{15}$&124.72 &65.33&51.76 & 73.57   \\
\tableline
  \end{tabular}
\tablecomments{The signs $a, b$ and $c$ denote: the values of $E_{F}(e), \langle E_{\nu} \rangle$\\
and $\langle E_{n}\rangle$ are calculated by  using  the relations of $E_{F}(e)\simeq$\\
43.44$(\frac{B}{B_{cr}})^{\frac{1}{4}}(\frac{\rho}{\rho_{0}}\frac{Y_{e}}{0.12})
^{\frac{1}{4}}$ MeV , $\langle E_\nu \rangle= \int E_\nu d\Gamma /\Gamma$and $\langle E_{n}\rangle = \\
E_{F}(e)- \langle E_{\nu}\rangle+0.61$  MeV, respectively.  We set $\rho=\rho_{0}$ and $Y_{e}\\
=0.12$, further details are presented in Appendix. }
\end{table}

\section{The calculation of magnetar soft X/$\gamma$-ray luminosity}
This section is composed of three subsections.  For each
subsection we present different methods and considerations.
\subsection{Physics on $L_{X}$ of a magnetar}
In this part, we  briefly  present a possible explanation for the
soft X/$\gamma$-ray luminosity of a magnetar.

As mentioned above, once the energy of electrons near the Fermi
surface are higher than the Fermi energy of neutrons ($E_{F}^{'}
(n) \approx$60 MeV, c.f. Shapiro \& Teukolsky 1983) the process of
EC will dominate.  Owing to superhigh density of the
star internal matter, the outgoing neutrons can't escape
from the star. In the interior of a NS, $\beta-$decay and
inverse $\beta-$decay always occurs simultaneously, as
required by charge neutrality \citep{Gamov41,Pethick92}. In the
'recycled' process: inverse $\beta-$decay $\rightarrow$
$\beta-$decay$\rightarrow$ inverse $\beta-$decay, the kinetic
energies of electrons resulting in the process of $\beta-$decay
are still high (higher than the neutron Fermi kinetic energy),
most of the electron energy loss is carried away by neutrinos
(antineutrinos) produced in this `recycle' process, only a small
fraction of this energy loss can effectively  contribute to heat
the star internal matter. If one outgoing neutron collide with
one ${}^3P_2$ Cooper pair, the ${}^3P_2$ Cooper pair with low
energy gap will be destroyed quickly.  The outgoing neutron will
react with the neutrons produced in the process $n + (n\uparrow
n\downarrow)\longrightarrow n+ n+ n$, the kinetic energy of
the outgoing neutrons will be transformed into thermal energy.
When accumulating to some extent, the transformed thermal energy
would transport from the star interior to the star surface by
conduction, then would be converted into radiation energy as
soft X-rays and $\gamma$-rays, $kT \simeq 10B_{15}$ keV. However,
most of thermal energy transported to the
star surface is carried away by the surface neutrino flux, only
a small fraction can be converted into radiation energy as soft
X/$\gamma$-rays.  After a highly efficient modulation within the
pulsar magnetosphere, the surface thermal emission(mainly soft
X/$\gamma$-rays) has been shaped into a spectrum with the observed
characteristics of magnetars.  It is worth noting that because of
the absorption of the star matter and the emission of neutrinos
escaping from the interior of a magnetar, the overwhelming majority
of the thermal energy will be lost in the process of energy
transportation. This lost thermal energy may maintain a relative
thermodynamic equilibrium in the interior of a magnetar.  The
energies of neutrinos escaping from the star interior could be
high as $kT\sim several$ MeV due to neutrinos' coherent scattering
caused by electrons(protons) and neutrons \citep{Shapiro83}, and the
heat carried away by neutrinos(antineutrinos) could be slightly
larger than that absorbed, despite of the compactness of the star
matter. Therefore, the whole electron capture reaction process
(here we focus on the direct Urca process) can be seen as a long-term
process of magnetic field decay, accompanied by magnetar's inner
cooling.  In a magnetar $L_{X}$ is ultimately determined by magnetic
field strength $B$, and is a weak function of the internal temperature
$T$. It should be noted that the surface temperature $T_{BB}$ is
controlled by crustal physics, and is independent of the evolution of
the core, while the internal temperature is only equivalent to
background temperature and decreases with decreasing $B$.
\subsection{The Calculation of $L_{X}$}

Actually, only the neutrons lying in the vicinity of the Fermi surface
are capable of escaping from the Fermi sea.  In other words, for degenerate
neutrons, only a fraction $\sim kT/E_{F}^{'}(n)$ can effectively contribute
to the electron capture rate $\Gamma$.  The rate of the total
thermal energy released in the EC process is calculated by
\begin{eqnarray}
&&\frac{dE}{dt} \simeq \frac{(kT)^{4}exp(-\Delta_{max}({}^3P_2)/kT)}{E_{F}^{'}(n)E_{F}^{'}(p)E_{F}(e)E_{F}(\nu_{e})}V({}^3P_2)\nonumber\\
 &&\times \frac{(2\pi)^{4}}{\hbar V_1}G_{F}^{2}C_{V}^{2}(1+3a^{2}) \int d^{3}n_{e}d^{3}n_{p}d^{3}n_{n}d^{3}n_{\nu}\nonumber\\
 &&\times\delta(E_{\nu}+Q-E_{e}) \delta^{3}(\overrightarrow{K_{f}}- \overrightarrow{K_{i}})S\langle E_{n}\rangle,
\end{eqnarray}
where $V({}^3P_2)$ denotes the volume of ${}^3P_2$ anisotropic neutron
superfluid ($V({}^3P_2) = \frac{4}{3}\pi R_{5}^{3})$; $V_1$ is the
normalized volume; $S = f_{e}f_{p}(1-f_{n})(1-f_{\nu})$, $f(j)
=[exp((E_{j}-\mu_{j})/kT)+ 1]^{-1}$ is the fraction of phase space
occupied at energy $E_{j}$ (Fermi-Dirac distribution), factors of
$(1-f_{j})$ reduce the reaction rate, and are called `blocking factor',
each factor of $d^{3}n_{j}$ must be multiplied by $(1-f_{j})$, in the
interior of a NS, for neutrinos (antineutrinos), $(1-f_{\nu})$ = 1;
for electrons, when $E_{e} < E_{F}(e)$, $f_{e}$ = 1, when $E_{e} >
E_{F}(e)$, $f_{e}$ = 0; for neutrons, when $E_{k}(n) < E_{F}(n)$,
$(1-f_{n})$ = 0, when $E_{k}(n) > E_{F}(n)$, $(1-f_{n})$ = 1; for
protons: when $E_{p} < E_{F}(p)$, $f_{p}$= 1, when $E_{p} > E_{F}(p)$,
$f_{p}$ = 0, so $S = f_{e}f_{p}(1-f_{n})(1-f_{\nu})\simeq 1$ can be
ignored in the latter calculations; the 4 powers of $kT$ originate
as follows: The reaction $e^{-}+ p \leftrightarrow n+ \nu_e$ in
equilibrium gives $dY_{e}=dY_{p}=-dY_{n}=-dY_{\nu_{e}}$, where $Y_{i}$
is the concentration of the $i$th species of particle \citep{Shapiro83}; in addition to
this, for each degenerate species, only a fraction $\sim \frac{kT}
{E_{F}(i)}$ can effectively contribute $\Gamma$, both $\Gamma$ and
$ L_{X}$ are proportional to $(kT)^{4}$.  In the interior of a NS,
the neutrons are 'locked' in a superfluid state, the rates for all
the reactions including $\beta$-decay and inverse $\beta$-decay are
cut down by a factor $\sim exp(-\Delta_{max}({}^3P_2)/kT)$, where
$\Delta_{max}({}^3P_2)\sim$ 0.048 MeV, is the superfluid energy gap
\citep{Elgar96}.  For convenience, we use the symbol $\Lambda$,
called `Landau level-superfluid modified factor', to represent $\frac{(kT)
^{4}exp(-\Delta_{max}({}^3P_2)/kT)}{E_{F}^{'}(n)E_{F}^{'}(p)E_{F}(e)
E_{F}(\nu_{e})}$.  The whole EC process (or the process of decay
of magnetic fields) can be seen as a long-term process of the inner
temperature's fall.  Due to the obvious effect of restraining direct
Urca reactions by neutron superfluid, the process of magnetar cooling
and magnetic field decay proceeds very slowly, so the value of `Landau
level-superfluid modified factor' $\Lambda$ can be treated as a
constant in the latter calculations. Since magnetars are different,
their initial reaction conditions (such as $B$, $T$, etc) are also
different. However, for simplicity, we assume the initial magnetic
field strength $B_{0}$ to be $3.0 \times 10^{15}$ G for all magnetars
in this work. The energy gap maximum of ${}^3P_2$ is $\Delta_{max}({}^3P_2)
\sim$ 0.048 MeV \citep{Elgar96}, the critical temperature of the
${}^3P_2$ neutron Cooper pairs can be evaluated as follows: $T_{cn} =
\Delta_{max}({}^3P_2)/2k \simeq $2.78 $\times 10^{8}$ K, so the maximum
of the initial internal temperature $T_{0}$ (not including the inner core)
can not exceed $T_{cn}$\citep{Peng06}.  Keeping $\Lambda$ as a constant, we
numerically simulate the process of magnetar cooling and magnetic
field decay. The details are shown in Table 2.

\begin{table}
\small
\caption{Numerical simulating magnetar cooling and magnetic field decay.  \label{tb1-5}}
\begin{tabular}{@{}crrrrr@{}}
\tableline
$B$      & \multicolumn{1}{c}{$T$\tablenotemark{1}}  &  \multicolumn{1}{c}{$T$\tablenotemark{2}} & \multicolumn{1}{c}{$T$\tablenotemark{3}} & \multicolumn{1}{c}{$T$\tablenotemark{4}}& \multicolumn{1}{c}{$T$\tablenotemark{5}}\\
 (G) & ($10^{8}$K)& ($10^{8}$K)& ($10^{8}$K)& ($10^{8}$K)& ($10^{8}$K) \\
\tableline
3.0$\times10^{15}$& 2.70  &2.65 &2.60  &2.55 &2.50\\
2.8$\times10^{15}$&  2.68 &2.63 &2.58  &2.53 &2.48 \\
2.5$\times10^{15}$&  2.64 &2.59 &2.54  &2.50&2.45 \\
2.0$\times10^{15}$&  2.57 &2.52&2.48  &2.43&2.38\\
1.5$\times10^{15}$&  2.48 &2.43 &2.39  &2.34&2.30  \\
1.0$\times10^{15}$& 2.34 &2.30 &2.26  &2.22&2.18   \\
9.0$\times10^{14}$& 2.30 &2.26 &2.22  &2.18&2.14\\
7.0$\times10^{14}$&  2.22 &2.18 &2.14  &2.10 &2.06 \\
5.0$\times10^{14}$& 2.09 &2.06 &2.02  &1.99    &1.95   \\
4.0$\times10^{14}$ & 2.00 &1.97 &1.94 &1.90&1.87   \\
3.0$\times10^{14}$& 1.87 &1.84 &1.81  &1.77&1.74\\
2.5$\times10^{14}$&  1.77 &1.74 &1.71  &1.69&1.66 \\
2.0$\times10^{14}$& 1.61 &1.58 &1.56  &1.54&1.51\\
1.6$\times10^{14}$&  1.23&1.23 &1.21  &1.19&1.18\\
\tableline
\tableline
  \end{tabular}
\tablecomments{The signs $1, 2, 3, 4, 5 $ denote
that the values of $\Lambda$ are 4.031\\
$\times10^{-14}$, 3.598$\times10^{-14}$, 3.202 $\times10^{-14}$, 2.841 $\times10^{-14}$ and 2.512$\times$\\
$10^{-14}$  corresponding to column 2, column 3, column 4, column 5 \\
and column 6, respectively.}
\end{table}

From the simulations above, we infer that, the magnetic field strength
$B$ is a weak function of the internal temperature $T$, which is only
equivalent to background temperature and decreases with decreasing $B$.
From Table 2, the mean value of $ \Lambda $ is 3.237 $\times
10^{-14}$. According to our model, the observed soft X/$\gamma$-ray output
of a magnetar is dominated by the transport of the magnetic field energy
through the core.  In order to obtain $L_{X}$, we must introduce
another important parameter $\zeta$, called ` effective X/$\gamma$-ray
coefficient' of a magnetar.  The main reasons for introducing
$\zeta$ are presented as follows:
\begin{enumerate}
\item Firstly, the thermal energy transported to the surface of a
magnetar could not be converted into the electromagnetic radiation
energy entirely (to see Section 1), therefore, we introduce an energy
conversion efficiency, $\epsilon$, which is defined as the ratio of
the amount of the soft X/$\gamma$-ray radiation energy converted to
the amount of the thermal energy transported to the star surface by
heat conduction.

\item Second, in the process of heat conduction, the lost thermal
energy could be either absorbed by the star inner matter, or carried away
by neutrinos. Thus, we introduce the thermal energy transfer coefficient
$\theta$, defined as the ratio of the amount of net thermal energy transported
to the star surface to the amount of the total thermal energy converted by
the magnetic field energy.

\item Finally,  we define the effective soft X/$\gamma$-ray coefficient of a magnetr
$\zeta$, $\zeta= \epsilon \theta$.  Due to special circumstances inside NSs
(high temperatures, high-density matter and superhigh magnetic fields etc),
the calculations of $\epsilon$ and $\theta$ have not yet appeared so far in
physics community, so it is very difficult to gain the values of $\epsilon$
and $\theta$ directly. However, the mean value of $\zeta$ of magnetars can be estimated
roughly by comparing  the calculations with the observations in our model.
Furthermore, we make an assumption that $L_{X}\propto \langle\zeta\rangle$ in
a magnetar.
\end{enumerate}

By introducing parameters $\zeta$ and $\Lambda$, magnetar soft
X/$\gamma$-ray luminosity $L_{X}$ can be computed by
\begin{equation}
L_{X}=  \langle\zeta \rangle  \frac{dE}{dt}.
\end{equation}

Inserting $\rho_{j}dE_{j}=d^{3}n_{j}$ and $\delta^{3}(\overrightarrow
{K_{f}}- \overrightarrow{K_{i}})d^{3}n_{p}(2\pi)^{3}/V_{1}$
=$\delta^{3}(\overrightarrow{P_{f}}-\overrightarrow{P_{i}})
d^{3}P_{p}$ into Eq.(16) gives
\begin{eqnarray}
 &&L_{X} \simeq \langle \Lambda \rangle \langle\zeta \rangle V({}^3P_2)\frac{2\pi}{\hbar}G_{F}^{2}C_{V}^{2}\nonumber\\
 && \times (1+3a^{2})\int \rho_{e}dE_{e} \int \delta^{3}(\overrightarrow{P_{f}}- \overrightarrow{P_{i}})d^{3}P_{p}    \nonumber\\
&&\times \int \langle E_{n}\rangle \rho_{n}dE_{n}\int \delta(E_{\nu}+Q-E_{e})\rho_{\nu}dE_{\nu}.
\end{eqnarray}
Eliminating $\delta$-functions and simplifying Eq.(17) gives
\begin{eqnarray}
&&L_{X}=\langle\Lambda\rangle \langle\zeta \rangle V({}^3P_2)\frac{2\pi}{\hbar}\frac{1}{2\pi^{2}\hbar^{3}c^{3}}G_{F}^{2}C_{V}^{2}(1+3a^{2}) \nonumber\\
&&\int_{E_{F}^{'}(n)}^{\langle E_{n}\rangle}\langle E_{n}\rangle \rho_{n}dE_{n}\int_{Q}^{E_{F}(e)}(E_{e}-Q)^{2}\rho_{e}dE_{e}.
\end{eqnarray}
Inserting Eq.(3) and Eq.(8) into Eq.(18) gives a general formula for $L_{X}$,
\begin{eqnarray}
&& L_{X} \simeq \langle \Lambda \rangle \langle\zeta \rangle \frac{4}{3}\pi R_{5}^{3}
 \frac{2\pi}{\hbar}\frac{G_{F}^{2}C_{V}^{2}(1+3a^{2})}{2\pi^{2}\hbar^{3}c^{3}}\frac{8\pi\sqrt{2}m_{n}^{\frac{3}{2}}}{h^{3}}\nonumber\\
&& \times \frac{(1.60 \times 10^{-6})^{8.5}}{2\pi^{2}\hbar^{3}c^{3}}
\int_{E_{F}^{'}(n)}^{\langle E_{n}\rangle}E_{n}^{\frac{1}{2}}\langle E_{n}\rangle dE_{n}\nonumber\\
&&\times \int_{Q}^{E_{F}(e)}(E_{e}^{2}-m^{2}_{e}c^{4})^{\frac{1}{2}}E_{e}(E_{e}-Q)^{3}dE_{e},
\end{eqnarray}
where the relation 1 MeV=1.6$\times 10^{-6}$ erg is used.
Inserting all the values of the following constants:$G_{F}$=1.4358
$\times 10^{-49}$ erg~cm$^{3}$, $C_{V}$=0.9737, $a$=1.253,
$\hbar$=1.05$\times 10^{-27}$ erg~s$^{-1}$, $h$=6.63$\times
10^{-27}$ erg~s$^{-1}$, $m_{e}$=9.109$\times 10^{-28}$ g,
$m_{n}$=1.67 $\times 10^{-24}$ g, $m_{e}c^{2}$=0.511 MeV and  $c$=3$\times 10^{10}$ cm~s
$^{-1}$ into Eq.(19) gives the values of $L_{X}$
in different superhigh magnetic fields.  Now, the calculations are partly listed as follows (to see Table 3).  
\begin{table}
\small
\caption{The  relation of  $L_{X}$ and $B$ \label{tb1-6}.}
\begin{tabular}{@{}crrrr@{}}
\tableline
 B       & $E_{F}(e)$  &   $\langle E_{n}\rangle$  & $L_{X}$\\
 (G)       &(MeV)          & (MeV)      &(erg~s$^{-1}$) &\\
\tableline
2.0$\times 10^{14}$&63.38    &60.98  & 3.924 $\times10^{32} \langle\zeta \rangle $ \\
4.0$\times 10^{14}$ &75.37    &63.50 & 5.051 $\times10^{35} \langle\zeta \rangle $ \\
5.0$\times 10^{14}$ &79.69   &64.67  & 1.973 $\times10^{36} \langle\zeta \rangle $\\
7.0$\times 10^{14}$&86.69   &66.15   & 1.010 $\times10^{37} \langle\zeta \rangle $  \\
9.0$\times 10^{14}$&92.31   &67.30   & 2.888 $\times10^{37} \langle\zeta \rangle $ \\
1.0$\times 10^{15}$&94.77    &67.80  & 4.350 $\times10^{37} \langle\zeta \rangle $ \\
1.5$\times 10^{15}$&104.88   & 69.79 & 1.849$\times10^{38} \langle\zeta \rangle $ \\
2.0$\times 10^{15}$&112.70   &71.30   & 4.701 $\times10^{38} \langle\zeta \rangle $ \\
2.5$\times 10^{15}$&119.17   &72.53    & 9.308 $\times10^{38} \langle\zeta \rangle $ \\
3.0$\times 10^{15}$&124.72  & 73.57  & 1.589$\times10^{39} \langle\zeta \rangle$ \\
\tableline
  \end{tabular}
\tablecomments{We assume that $\langle\Lambda \rangle$=
3.237$\times 10^{-14}$, $\rho=\rho_{0}$ \\
and $Y_{e}$=0.12 when calculating $L_{X}$. }
\end{table}

If we want to determine the value of $\langle\zeta \rangle$, we must
combine our calculations with the observed persistent parameters of
magnetars. The details are to see in  $\S$ 3.3 and $\S$ 3.4.
\subsection{ Observations of magnetars}

Up to now, nine SGRs (seven conformed) and
twelve AXPs (nine conformed) at hand, a statistical investigation
of their persistent parameters is possible.  Observationally, all
known magnetars are X-ray pulsars with luminosities of $L_{X}\sim
(10^{32}\sim 10^{36})$ erg~s$^{-1}$, usually much higher than the
rate at which the star loses its rotational energy through spin-
down \citep{Rea10}. In Table 4, the persistent parameters of
sixteen conformed magnetars are listed in the light of observations
performed in the last two decades.
\begin{table*}[t]
\small
\caption{The observational  parameters of  magnetars confirmed.  \label{tb1-7}.}
\begin{tabular}{@{}crrrrrr@{}}
\tableline
Name& $P$& $\dot{P}$& \multicolumn{1}{c}{$T_{BB}$\tablenotemark{b}}& \multicolumn{1}{c}{$B$\tablenotemark{c}}&$L_{X}$& \multicolumn{1}{c}{$dE/dt$\tablenotemark{e}}\\
 $$& (s) & ($10^{-11}$s ~s$^{-1}$)&($10^{6}$ K)& ($10^{14}$ G)& (erg~s$^{-1}$) & (erg~s$^{-1}$)\\
\tableline
SGR0526-66   & 8.0544 &3.8  &NO        & 5.6       & 1.4$\times10^{35}$       & 2.9$\times10^{33}$\\
SGR1806-20   & 7.6022 &75   &6.96        & 24        &5.0$^{a}\times10^{36}$    & 6.7$\times10^{34}$ \\
SGR1900+14   & 5.1998 &9.2   &5.45        &7.0       &(0.83$\sim$1.3)          &2.6$\times10^{34}$ \\
           &         &       &        &           &  $\times10^{35}$                             &        \\
SGR1627-41    & 2.5946 &1.9 &NO             & 2.2      & 2.5$\times 10^{33}$      & 4.3$\times 10^{34}$\\
SGR0501+4516 & 5.7621 &0.58  & 8.1         & 1.9        & NO                       & 1.2$\times 10^{33}$\\
SGR0418+5729 & 9.0784 &$<$    &No   &$<$  & NO                      &$<$         \\
              &       &0.0006 &    &0.075            &             & 3.2$\times10^{29}$\\
SGR1833+0832 & 7.5654 &0.439  & No          &1.8        & NO                       & 4.0$\times10^{32}$\\
\tableline
CXOUJ0100    &8.0203 &1.88     &4.41     & 3.9        & 7.8$\times10^{34}$      & 1.4$\times10^{33}$\\
1E2259+586 & 6.9789  &0.048    &4.77      & 0.59        & 1.8$\times10^{35}$      &5.6$\times10^{31}$\\
4U0142+61  & 8.6883  &0.196    &4.58      &1.3       & $>$5.3$\times10^{34}$    &1.2$\times10^{32}$\\
1E1841-045 &11.7751  &4.155     &5.10     &7.1       & 2.2$\times10^{35}$      &9.9$\times10^{32}$\\
1RXSJ1708  &10.9990  &1.945     &5.29    &4.7        & 1.9$\times10^{35}$      &5.7$\times10^{32}$\\
CXO$^{t}$J1647&10.6107 &0.24     &7.31    &1.6       & 2.6$\times10^{34}$  &7.8$\times10^{31}$\\
1E$^{t}$1547.0-5408 &2.0698  &2.318 &4.99 &2.2     & 5.8$\times10^{32}$      &1.0$\times10^{35}$\\
XTE$^{t}$J1810-197 &5.5404  &0.777  &1.67  &2.1     &1.9$\times10^{32}$      &1.8$\times10^{33}$\\
1E$^{d}$1048.1-5937&6.4521 &2.70   &7.23  &4.2      &5.4$\times10^{33}$      &3.9$\times10^{33}$\\
\tableline
  \end{tabular}
\tablecomments{All data are from the McGill
AXP/SGR online catalog of 10 April  2011(http://www.physics. mcgill.
ca/$^{\sim}$pulsar/magnetar/ main.html) except for $L_{x}$ of SGR 1806
-20. The sign $a$ denotes: from Thompson \& Duncan (1996). The sign $b$
denotes: the data of column 3 are gained from the original data by using
the approximate relation 1 keV $\sim$ 1.16$\times 10^{7}$ K ($T\sim E/k$,
$E$ and $k$ are the energy of a photon and the Boltzmann constant, respectively).
The sign $c$ denotes: The surface dipolar magnetic field of a pulsar can
be estimated using its spin period, $P$, and spin-down rate, $dot{P}$, by
$B\simeq 3.2\times 10^{19}(P \dot{P})^{\frac{1}{2}}$ G.  The signs $d$
and $t$ denote: dim AXP and transient AXP, respectively. The sign $e$ denotes:
A pulsar slow down with time as its rotational energy is lost via magnetic
dipolar radiation, and the loss rate of a pulsar's rotational energy is
noted as $dE/dt$.}
\end{table*}

From Table 4, three magnetars SGR0501+4516, SGR0418+5729 and
SGR1833+0832 with no persistent soft X/$\gamma$-ray fluxes
observed will not be considered in the latter calculations.  Although the lack of optical
identifications restrict accurate distance estimates for some
magnetars, it is clear from their collective properties (such as
high X-ray absorption and distribution in the Galactic plane, etc)
that these sources have characteristic distances of at least a few kpc.
Such values, supported in some cases by the distance estimates of the
associated SNRs, imply typical $L_{X}$ in the range $10^{34}\sim 10^{36}$
erg~s$^{-1}$, clearly larger than the rotational energy loss inferred from
their period and $\dot{P}$ values.  Moreover, according to the magnetar
model\cite{Duncan92,Duncan96,Thompson96}, the persistent soft X/$\gamma$-ray
luminosity of a canonic magnetar shouldn't be less than its  rotational
energy loss rate $dE/dt$.  In order to reduce the error in the
calculation of the average  value of $\zeta$, all the transient magnetars
listed in Table 4 (including SGR1627-41, CXOJ1647, 1E11547.0-5408 and  XTEJ
1810-197) are no longer to be considered in the latter calculations.
Combining Eq.(20) with Eq.(4) gives the values of $\zeta$ of
magnetars.  The calculations are shown as below (to see Table 5).
\begin{table}
\small
\caption{The values of $\zeta$ of magnetars  \label{tb1-8}.}
\begin{tabular}{@{}crrrr@{}}
\tableline
Name&$B$ &$E_{F}(e)$ &$\langle E_{n}\rangle $  &$ \zeta $\\
 $$& (G) & (MeV)& (MeV)& $$\\
\tableline
SGR0526-66   &5.6$\times10^{14}$ &81.98       &63.93     & 5.32$\times10^{-2}$       \\
SGR1806-20   & 2.4$\times10^{15}$ &117.96       &72.30          & 6.07$\times10^{-3}$ \\
SGR1900+14   & 7.0$\times10^{14}$ &86.69        &66.15         &1.28$\times10^{-2}$ \\
CXOUJ0100    &3.9$\times10^{14}$ &74.89     & 63.63        & 1.67$\times10^{-1}$\\
1E1841-045 &7.1$\times10^{14}$  &86.99    &66.21      & 2.05$\times10^{-2}$      \\
1RXSJ1708  &4.7$\times10^{14}$  &78.47     &64.41       & 1.34$\times10^{-1}$   \\
1E1048.1-5937&4.2$\times10^{14}$ &76.29  &63.93      &7.39$\times10^{-3}$     \\
\tableline
  \end{tabular}
\tablecomments{We assume that $\langle\Lambda \rangle$=
3.237$\times 10^{-14}$, $\rho=\rho_{0}$ and $Y_{e}$\\
= 0.12 when calculating $\zeta$.}
\end{table}

Clearly from Table 6, the values of $\zeta $ of  most magnetars
are about $10^{-1}\sim 10^{-3}$.  Since the value of $\zeta$ is
mainly determined by the magnetic field strength of a magnetar, the mean
value of $\zeta$ of magnetars can be roughly estimated by
\begin{equation}
 \langle\zeta \rangle = \frac{\sum B_{i}\zeta_{i}}{\sum B_{i}}.
\end{equation}
Employing  Eq.(20) gives the mean value of $ \zeta $ of magnetars $\sim$3.803
$\times 10^{-2}$.  Theoretically, using $\langle\zeta \rangle$ allows
us calculate the value of $L_{X}$ in any ultrastrong magnetic field. The details
are to see in $\S$ 3.4.

\subsection{Comparing calculations with observations}
Inserting $\langle\zeta \rangle$ into  Eq.(20) gives the value
of $L_{X}$ in any  strong magnetic field. Now, the calculations are
partly listed as follows.
\begin{table}
\small
\caption{The calculated values of  $L_{X}$ for magnetars \label{tb1-9}.}
\begin{tabular}{@{}crrrrr@{}}
\tableline
 B       & $E_{F}(e)$  &  $\langle E_{\nu}\rangle$  &  $\langle E_{n}\rangle$  & $L_{X}$\\
 (G)       &(MeV)         &(MeV) & (MeV)      &(erg~s$^{-1}$) &\\
\tableline
2.0$\times 10^{14}$&63.38 & 3.01  &60.98  & 1.492$\times10^{31} $ \\
5.0$\times 10^{14}$ &79.69 &15.63  &64.67  & 7.503 $\times10^{34}  $\\
1.0$\times 10^{15}$&94.77  & 27.58 &67.80  & 1.654 $\times10^{36}  $ \\
1.5$\times 10^{15}$&104.88 &35.70  & 69.79 & 7.032$\times10^{36}  $ \\
2.0$\times 10^{15}$&112.70 & 42.01 &71.30   & 1.788 $\times10^{37} $ \\
2.8$\times 10^{15}$&122.59 &50.03  &73.17    &4.948 $\times10^{37}  $ \\
\tableline
  \end{tabular}
\tablecomments{We assume that $\langle\Lambda \rangle$=
3.237$\times 10^{-14}$, $\langle\zeta \rangle$=3.083$\times10^{-2}$,\\
$\rho=\rho_{0}$ and $Y_{e}$=0.12 when calculating $L_{X}$. }
\end{table}
Furthermore, employing the mean value of $\zeta$, we also gain
the schematic diagrams of soft X/$\gamma$-ray luminosity as a
function of magnetic field strength.  The results of fitting
agree well with the observational results in three models,
which can be shown as in Figure 2.

 \begin{figure}[th]
\centering
  \vspace{0.5cm}
  \includegraphics[width=8.1cm,angle=360]{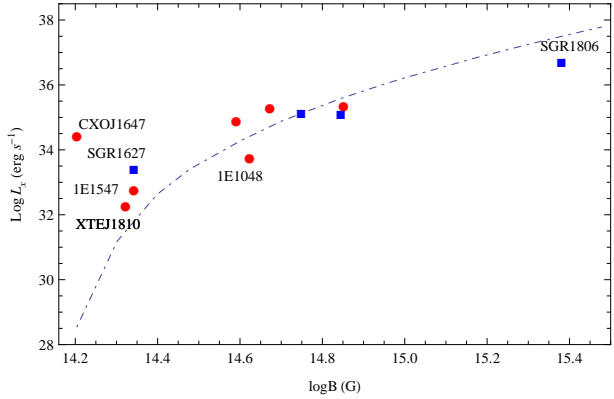}
  \caption{The diagram of soft X/$\gamma$-ray luminosity $L_{X}$
as a function of magnetic field strength $B$ when $Y_{e}$= 0.12
and $\rho =\rho_{0}$.  Circles and squares are for AXPs and SGRS,
respectively. The range of $B$ is assumed to be (1.60 $\times 10^{14}\sim $ 3.0$\times 10^{15}$ G)
considering that, when $B\leq B_{f}$, the direct Urca process ceases,
while the modified Urca process still occurs, from which weaker X-ray and
weaker neutrino flux are produced.}
  \label{4:fig}
\end{figure}

Clearly from Figure 2, the magnetar soft X/$\gamma$-ray luminosity
increases with increasing magnetic field obviously, and the
steepness of every fitting curve corresponding to weaker magnetic
fields ($B< 2.5 \times 10^{14}$ G) is larger than that corresponding
to higher magnetic fields $B\geq 2.5 \times 10^{14}$ G), because
Eq.(1) is  approximately hold only when $B\gg B_{cr}$.  For SGR 1806-20
with the lowest value of $\zeta$, whose  soft X/$\gamma$
-ray luminosity is cited from \citep{Thompson96}, its observed value of $L_{X}$ may be
biased by the intense low-frequency absorption, corresponding
to an electron column density of $\sim$ 6 $\times 10^{22}$
cm$^{-2}$ \citep{Murakami94,Sonobe94}, as a result, a significant
blackbody component contained in the X-ray bolometric flux is
undetected, the luminosity in relativistic particles needed to
power the plerion is $\sim$ 5 $\times 10^{36}\times( D/8{\rm
kpc})^{2.5}$ erg~s$^{-1}$ \citep{Thompson96}.  With respect to 1E
1048-59, which is discovered as a 6.4 s dim isolated pulsar near the
Carina Nebula \citep{Steward86}, substantial data were subsequently
obtained, showing unambiguous evidence for a large flux increase coupled
to a decrease in the pulsed fraction \citep{Mereghetti04}, as well
exemplified by the fact that its $L_{X}$ decreased from (1$\sim$2)$\times
10^{34}$ erg~s$^{-1}$ \citep{Mereghetti04} to 5.4 $\times 10^{33}$ erg~s$^{-1}$
between September 2004  and April 2011. Therefore, for SGR 1806-20 and 1E1048-59,
their values of $\zeta$ are lower than the mean value of $\zeta$ of magnetars,
which can be shown in Figure 2. For transient magnetars SGR1627-41, CXOJ1647, 1E11547
.0-5408 and  XTEJ1810-197, their observed soft X/$\gamma$-ray luminosities are high
than those calculated in theory (far from the fitted curve), the possible explanations are presented as follows:
\begin{enumerate}
\item Firstly, with respect to SGR1627-41,
its abnormal behavior suggests a connection between the bursting activity
and the luminosity of transient magnetars: in 1998  more than 100
bursts in about 6 weeks were observed with different satellites
\citep{Woods04}, however, no other bursts have been reported since
then. Its soft X-ray counterpart was identified with BeppoSAX in
1998 at a $L_{X}$ level of $10^{35}$ erg s$^{-1}$. Observations
carried out in the following 13 years showed a monotonic decrease
in its soft X/$\gamma$-ray luminosity, down to the current level of $\sim 2.5\times
10^{33}$ erg s$^{-1}$ (to see in Table 4).

\item Second,  the first transient AXP XTE J1810-197, discovered in 2003
\citep{Ibrahim04}, displaced a persistent flux enhancement by a factor of
$>$ 100 with respect to quiescent luminosity level $L_{X}$ increased from
7$\times 10^{32}$ erg~s$^{-1}$ to 5$\times 10^{34}$ erg~s$^{-1}$ between
June 1992 and September 2004 \citep{Bernardini09,Gotthelf04}, however, the
latest observations showed an obvious decline in $L_{X}$ (in 10 April 2011
$L_{X}$ $\sim$ 1.9 $\times 10^{32}$ erg~s$^{-1}$, to see in Table 6).
Several other transient magnetars (SGR1627-41,CXOJ1647 and 1E1547.0-5408)
have been discovered after XTEJ1810-197.  When shining, they have spectral and
timing properties analogous with those of the persistent sources.  During their
`quiescent' phases they universally possess luminosity of $\sim 10^{32}$ erg~s
$^{-1}$ and soft thermal spectra, that make them similar to CCOs
\citep{Mereghetti10}.  The long-term variations in $L_{X}$ of transient magnetars
could be associated with the bursting activities.  The source high state
coincided with a period of strong bursting activity, while in the
following years, during which no bursts were emitted, its luminosity
decreased \citep{Mereghetti08}.

\item. Finally, the magnetic field strengthes of these four transient magnetars
are in the range of $(1.6\sim 2.2 )\times 10^{14}$ G, and their values
of $L_{X}$ are calculated to be $\sim 10^{28}\sim 10^{31}$ erg~s$^{-1}$
according to our model.  It is not strange that the observed value
of $L_{X}$ of a transient magnetar is higher than that calculated if we
take into account of the long-term effect of the bursting activity on
the soft X/$\gamma$-ray luminosity of a transient magnetar.
\end{enumerate}

What must be emphasised here is that, for AXPs 4U 0142+61 and 1E 2259+586 their
mechanisms for persistent soft X/$\gamma$-ray may be related with the accretion, and
will be beyond of the scope of our model. The possible explanations are also
presented as follows:

\begin{enumerate}
\item Firstly, for AXPs 4U 0142+61 and 1E 2259+586 their magnetic field
strengthes are lower than the critical magnetic field $B_{f}$.  According
to our model, once $B\leq B_{f}$, the direct Urca process ceases, while the
 modified Urca process still occurs, from which weaker X-ray and
weaker neutrino flux are produced.

\item   Second, the observed properties of 1E 2259+586 seem consistent
with the suggestion that it is an isolated pulsar undergoing a
combination of spherical and disk accretion \citep{White84}. This
magnetar could be powered by accretion from the remnant of Thorne
-$\dot{Z}$ytow object(T$\dot{Z}$) \citep{van95}.

\item Finally, as concerns AXP 4U 0142+61, which was previously
considered to be a possible black hole candidate on the basis of its
ultra-soft spectrum \citep{White84}, the simplest explanation for its
$L_{X}$ involves a low-mass X-ray binary  with a very faint
companion, similar to 4U 1627-67\citep{Israel94}.
\end{enumerate}

Furthermore, timing observations show that the period derivative
of SGR 0418+5729, $\dot{P} <$ 6.0 $\times 10^{-15}$ s~s$^{-1}$,
which implies that the corresponding limit on the surface dipolar
magnetic field of SGR 0418+5729 is $B < $7.5 $\times 10^{12}$ G
\citep{Rea10}.  If the observations are reliable, then the value
of $L_{X}$ of SGR 0418+5729 calculated in our model will be far less
than $L_{X}\sim (10^{32}-10^{36})$ erg~s$^{-1}$, which implies that
our model is not in contradiction with the observation of SGR 0418+
5729 (the real value of $L_{X}$ of SGR 0418+5729 is too low to be observed so far).
SGR 0418+5729 may present a new challenge to the currently existing
magnetar models.  However, in accordance with the traditional view on the electron
Fermi energy, the electron capture rate $\Gamma$ will  decrease
with increasing $B$ in ultrastrong magnetic fields.  If the electron
captures induced by field-decay are an important mechanism powering
magnetar's soft X-ray emission \citep{Cooper10}, then $L_{X}$ will also
decrease with increasing $B$, which is contrary to the observed data
in Table 4 and the fitting result of Figure 2.

\section{Conclusions}
In this paper, by introducing two important parameters: Landau level-superfluid modified factor
and effective X/$\gamma$-ray coefficient, we numerically simulate the
process of magnetar cooling and magnetic field decay, and then  compute
$L_{X}$ of magnetars. We also present a necessary discussion after comparing
the observations with the calculations.  From the analysis and the calculations
above, the main conclusions are as follows:

1. In the interior of a magnetar, superhigh magnetic fields give rise
to the increase of the electron Fermi energy, which will induce electron
capture reaction.

2. The resulting high-energy  neutrons will destroy anisotropic ${}^3P_2$
neutron Cooper pairs, then the ${}^3P_2$ anisotropic superfluid and the
superhigh magnetic field induced by the ${}^3P_2$ Cooper pairs will
disappear.

3.By colliding, the kinetic energy of the outgoing neutrons will be
transformed into thermal energy. This transformed
thermal energy would be transported from the star interior to the star
surface by conduction, then would be converted into radiation energy
as soft X-rays and $\gamma$-rays.

4.The largest advantage of our models is not only to
explain but also to calculate magnetar soft X/$\gamma$-ray
luminosity $L_{X}$; further, employing the mean value of
$\zeta$, we obtain the schematic diagram of $L_{X}$ as a
function of $B$. The result of fitting agrees well with the
observation result.

Finally, we are hopeful that our assumptions and numerical
simulations can be combined with observations in the future,
to provide a deeper understanding of the nature of soft X/
$\gamma$-ray of a magnetar.

\begin{acknowledgments}
We are very grateful to Prof. Qiu-He Peng and Prof. Zi-Gao Dai
for their help in improving our presentation.   This work is supported
by National Basic Research Program of China (973 Program 2009CB824800),
Knowledge Innovation Program of The Chinese Academy Sciences KJCX$_{2}$
-YW-T09, Xinjiang Natural Science Foundation No.2009211B35, the Key
Directional Project of CAS and NSFC under projects 10173020,10673021,
10773005, 10778631 and 10903019.
\end{acknowledgments}

\appendix
\textbf{Appendix}
\section{ The effect of a superhigh magnetic field on $Y_{p}$ }
In the case of field-free, for reactions $e^{-}
+p \rightarrow n+ \nu_{e}$ and $n \rightarrow p +e^{-}+\nu^
{-}_{e}$ to take place, there exists the following inequality
among the Fermi momenta of the proton($p_{F}$), the electron
($k_{F}$)and the neutron ($q_{F}$): $p_{F}+ k_{F}\geq q_{F}$
 Together with the charge neutrality
condition, the above inequality brings about the threshold
for proton concentration $Y_{p}\geq $1/9, this means that,
in the field-free case, direct Urca reactions are strongly
suppressed by Pauli blocking in a system composed of neutrons,
protons, and electrons .  In the core of a NS,
where $\rho \geq 10^{15}$ g~cm$^{-1}$ and $Y_{p}$ could be
higher than 0.11, direct Urca processes could take place
\citep{Baiko99, Lai91, Yakovlev01}.  However, when in a superhigh
magnetic field $B\gg B_{cr}$, things could be quite different
if we take into account of the effect of the superhigh
magnetic field on $Y_{p}$.  The effect of a superhigh
magnetic field on the NS profiles is gained by applying
equations of state (EOS) to solve the Tolman-Oppenheimer-Volkoff
equation \citep{Shapiro83}.  In the paper of Chakrabarty S. et al
(1997), employing a relativistic Hartree theory, authors
investigated the gross properties of cold symmetric matter
and matter in $\beta-$equilibrium under the influences of
ultrastrong magnetic fields. The main conclusions of the paper
include: There could be an extremely intense magnetic field $\sim
10^{20}$ G inside a NS; $Y_{p}$ is a strong function of $B$ and
$\rho$; when $B$ is near to $B_{cr}^{p}$, the value of $Y_{p}$
is expected to be considerably enhanced, where $B_{cr}^{p}$ is
the quantum critical magnetic field of protons ($\sim$1.48
$\times10^{20}$ G); by strongly modifying the phase spaces of
protons and electrons, magnetic fields of such magnitude ($\sim
10^{20}$ G) can cause a substantial $n\rightarrow p$ conversion,
as a result, the system is converted to highly proton-rich matter
with distinctively softer EOS, compared to the field-free case.
Though magnetic fields of such magnitude inside NSs are
unauthentic, and are not consistent with our model ($B\sim
10^{14-15}$ G), their calculations are useful in supporting
our following assumptions: when $B\sim 10^{14-15}$ G, the value
of $Y_{p}$ may be enhanced, and could be  higher than
the mean value of $Y_{p}$ inside a NS ($\sim$ 0.05); direct Urca reactions are expected to occur inside a magnetar.  Based
on these assumptions, we can gain a concise expression for $E_{F}(e)$,
$E_{F}(e)\simeq$43.44$(\frac{B}{B_{cr}})^{\frac{1}{4}}(\frac{\rho}{\rho_{0}}\frac{Y_{p}}{0.12})
^{\frac{1}{4}}$ MeV by solving Eq.(1) of this paper.
(Cited from Gao et al.(2011c))
\end{document}